# A Guidewire Trajectory Generation Tool for Interventional MRI Simulations

Fatima Ben Haj[1], Jeffrey Hand[3], Shaihan Malik[3], Felipe Godinez[2,4]

[1]Princeton University, New Jersey, Princeton, USA
[2]Department of Radiology, University of California Davis, Sacramento, California, USA
[3]School of Biomedical Engineering and Imaging Sciences, King's College London, London, UK
[4]Comprehensive Cancer Center, University of California Davis, Sacramento, California, USA

Corresponding Author: fgodinez@health.ucdavis.edu

## Abstract

**Significance:** Cancer therapies based on selective drug delivery in brain tumors using endovascular devices under MRI guidance have become the center of ongoing investigations. In addition to questions regarding impact on clinical practice, endovascular device safety in MRI remains a major limiting factor.
**Purpose:** To present a computational tool that rapidly generates large populations of randomized guidewire trajectories for studying radiofrequency (RF)-induced heating risk in interventional MRI, and to illustrate its use comparing heating risk between local and body transmit coils.
**Methods:** A guidewire trajectory generation (GTG) tool was developed in MATLAB that models a guidewire as a spline defined by user-specified control nodes and coordinate ranges, generating thousands of randomized three-dimensional trajectories of a prescribed nominal length. Trajectory length accuracy and runtime were characterized across spline interpolation densities. As a use case, RF electric fields for a local coil array and a body birdcage coil (both 64 MHz, matched B1+) were simulated in a Duke phantom. Tip heating was estimated for 1000 randomized trajectories per coil by combining GTG geometries with an electric transfer function calculated for a straight wire at multiple insertion depths, yielding a pseudo specific absorption rate (sSAR) per trajectory.
**Results:** With the default 500 interpolated points, trajectory length accuracy was 99.84% (99.98% at 3000 points), at an average runtime of 11.5 seconds for each trajectory. The local coil array produced 56% lower scattered field power at the guidewire tip on average than the body coil; body-to-local coil ratios at the 50th, 95th, and 99th percentiles were 1.5, 1.6, and 2.27, respectively. A subset of trajectories, however, showed a higher scattered field for the local coil.
**Conclusion:** The GTG tool enables large-scale, statistically powered characterization of guidewire heating risk without repeated full-wave simulation or phantom experiments. While local coils reduced heating risk on average, coil type alone was not a reliable predictor of risk for individual trajectories, underscoring the value of trajectory-level safety assessment.

## 1. Introduction

Cancer therapies based on selective drug delivery in brain tumors using endovascular devices under MRI guidance have become the center of ongoing investigations [1], [2], [3]. In addition to questions regarding impact on clinical practice, endovascular device safety in MRI remains a major limiting factor. In interventional MRI, the wide use of metallic endovascular devices - such as guidewires and catheters - is precluded by heating effects driven by the coupling of the endovascular device and the scanner's transmit radio frequency (RF) coils. This phenomenon has been extensively investigated using physical experimentation [4], which may not capture the wide variability of endovascular device geometry and placement in the human body. Alternatively, simulations are adaptable and computationally robust enough to capture the variability of guidewire trajectories [5], [6], [7]. In simulation, a large number of guidewire trajectories can be modelled efficiently, providing a less expensive and complementary approach to studying guidewire heating in MRI. Such simulations not only inform on the MRI wire heating phenomena, but they can also guide and fine-tune physical experimentation.

The geometry of a placed guidewire plays a vital role in the RF-coil/endovascular device interactions because it determines the possible incident angles of the RF field, whose angular dependence can couple either strongly or weakly to the guidewire, hence modulating the amount of electric scatter at the tip [8]. Typically, guidewire tip heating patterns are calculated using a 3D digital anthropomorphic model that mimics the dielectric properties and shapes of human tissue, and the dominant factors that affect heating variations [9]. A well-designed simulation will include guidewire length, guidewire placement/position, incident field strength/frequency, and guidewire insertion depth [10]. As shown by Konings et al.[10], guidewire insertion depth is a large factor in the amount of electric field scatter or heating produced at the guidewire tip during interventional MRI.

Previous research [5], [11] has assessed the external validity of simulation techniques that use digital phantoms and prescribed guidewire trajectories, reporting a relatively high simulation-to-experimental accuracy, with an $r^2$ value of 0.74, and the largest error in wire tip temperature less than 1 degree Celsius. That work modelled a small number (7 trajectories) of implanted wire trajectories inside a dielectric digital phantom using an electromagnetic (EM) simulator, which is highly labor-intensive. Because simulating a large spread of wire geometries in an EM simulator is challenging and time-consuming, we propose a new rapid analysis approach based on the wire tip heating transfer function (TF) model [12]. In our approach, the guidewire trajectories are rapidly generated with a numerical model and analyzed using TFs in the presence of an incident MRI RF field generated in an EM simulator. The TF is generated once in the EM simulator for a simple straight wire geometry and for the desired insertion depths. The goal is to create a simulation that allows users to quickly generate thousands of trajectories, creating data sets with more statistical power, capturing worst-case scenarios and dangerous outlier guidewire geometries that produce high heating at the tip.

In this work, we present the design and implementation of a guidewire trajectory modelling tool that rapidly generates random trajectories within user-defined geometrical ranges. We showcase the tool by performing a simulated comparison of guidewire heating effects between a local coil and a body coil [13], [14]. We demonstrate its use and characterize its performance with respect to runtime and nominal wire-length accuracy.

## 2. Methods

We propose a new guidewire simulation paradigm that has three core parts: (1) a large number of random guidewire trajectories, (2) calculated TFs for a simple geometry and insertion depths, (3) an RF field derived for a specific scanner and transmission coils. Parts (1) and (2) can be effectively simulated using an EM simulator. However, the former part poses a severe bottleneck in implanted wire simulations. We developed a tool with the objective of modeling the trajectory of a guidewire's long axis and generating many random trajectories with specified lengths. The guidewire trajectory generation (GTG) tool was built in MATLAB (R2022, MathWorks Inc., USA). The repository of the code with the tool implementation is available for open sharing on GitHub (https://github.com/mushrooomshinobi/wire_function). The GitHub repository contains more detailed descriptions of the variables and a run example.

*Transmit field generation*

Two incident RF electric fields were generated using an EM simulator software (Sim4life, Zurich, Switzerland), one with a volume coil and another with a multi-channel local coil. A full human-sized phantom, Duke 14, was excited with a local coil array with 8 coil loops (200mm x 50mm) in circularly polarized mode, with 4 coils posterior and 4 anterior. The same Duke phantom was also excited with a body-sized birdcage coil with 16 rungs, 45cm long, in a high-pass configuration. The electric field was scaled so that the B1 + field was approximately equivalent within a region of interest (10cm x 10cm x 10cm). Both coils were tuned to 64MHz, corresponding to 1.5T field strength MRI scanners. The details for generating the electric field map are outside the scope of this work and can be seen here[15].

*GTG model and function*

In this section, we describe the GTG main functions and variables. The GTG guidewire is modelled as a spline defined by a set of fixed node locations that are user inputs (fig. 1). The user can also define the coordinate range of each spline node, specifying in the GUI which nodes can randomly move (fig. 2). Moving node selection is done via a window displaying a 2D wire projection overlaid on the EM-simulated electric field distribution.

The list of GTG functions and variable descriptions can be found in Tables 1 and 2. The main spline generation function is called the "wire_function". It has four required inputs and seven optional inputs. The required inputs provide essential information about the initial position of the guidewire control nodes, the range the nodes are allowed to vary, and the electric field map to be referenced. Optional inputs enable visualization tools to monitor the output guidewire trajectories, including plotting the trajectory and histograms of the guidewire lengths.

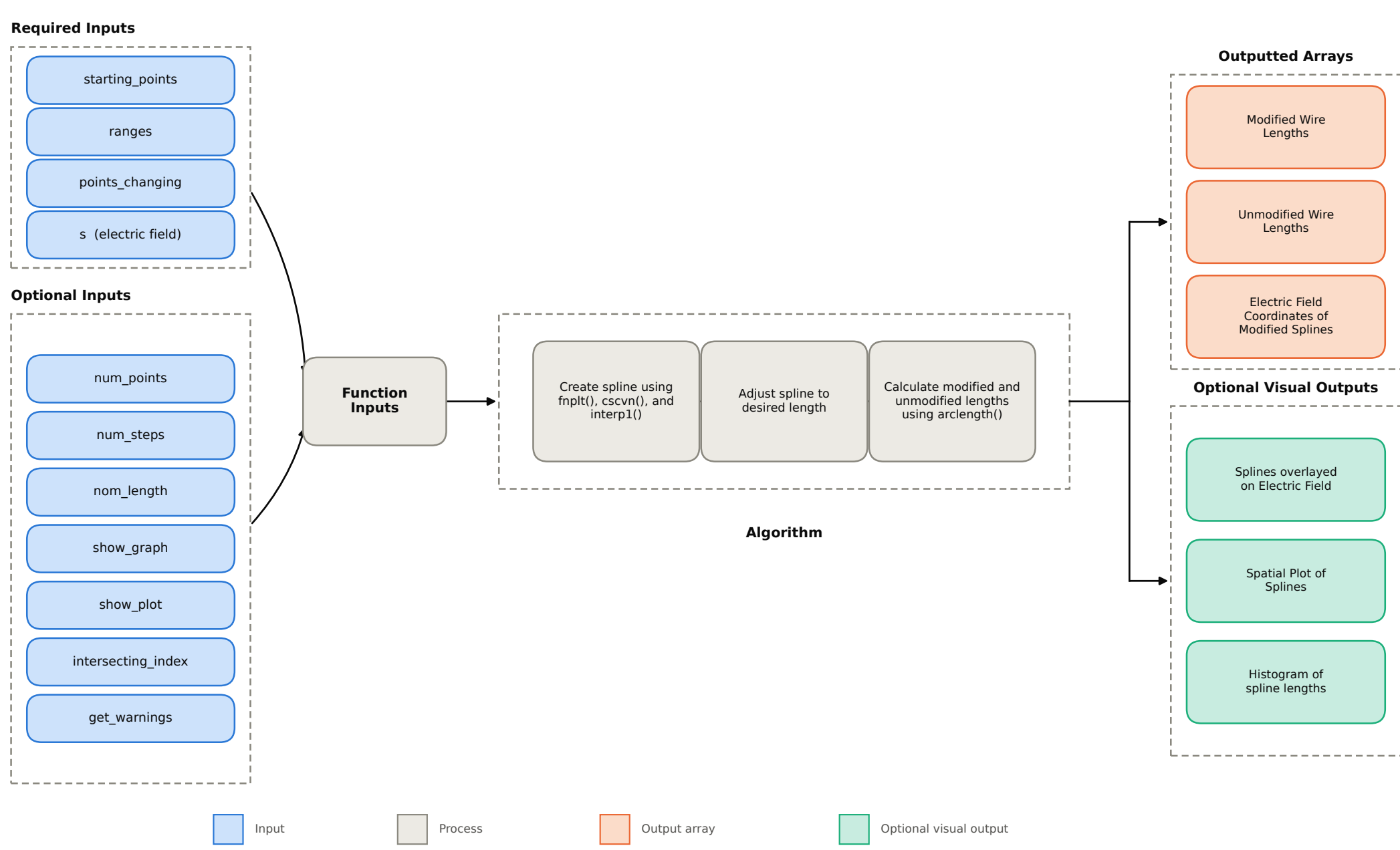

Figure. 1 - Block diagram of the guidewire trajectory generation tool (wire_function()). The user inputs the spline nodes (starting_points) and the range through which they can randomly move (ranges), as well as a variety of required and optional inputs. In the algorithm, the spline function connects the node points and generates a 3-dimensional path, which has a length that can be adjusted using the variable nom_length. This path is displayed as a 2D plot of the guidewire trajectory projection overlayed on a simulated electric field distribution.

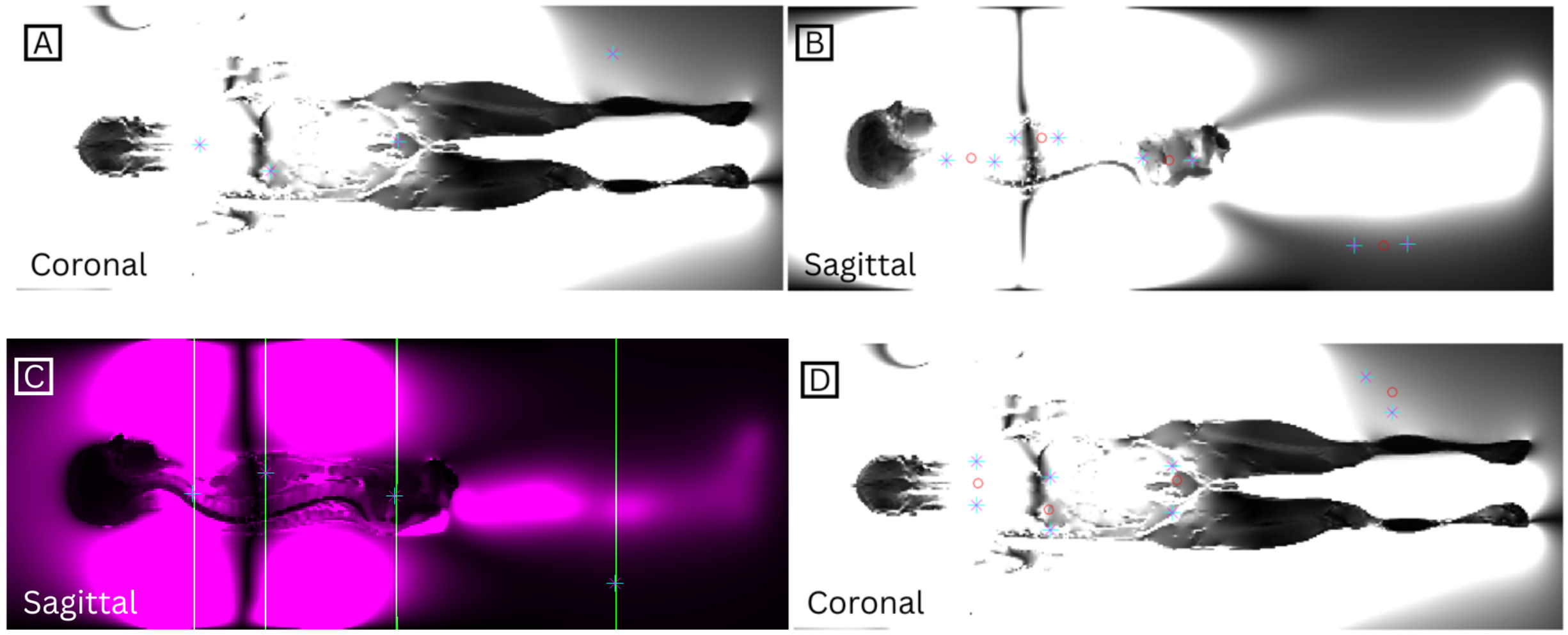

Figure. 2 – Snapshots from the Graphical User Interface in which the user can pick the starting points (A in the coronal plane and C in the sagittal plane) and ranges (B in the sagittal plane and D in the coronal plane) to define their spline. The magenta and blue x's appear on the image as points are selected using MATLAB's getpts function. The red circles and green lines represent locations where points have already been selected.

Table 1. Functions defined. The function arclength() was adopted from MATLAB central file exchange [16] .

| | Function | Syntax Example |
|---|---|---|
| MATLAB library functions | cscvn()<br>- Constructs the spline<br>fnplt()<br>- returns the matrix of points that would be plotted from the cscvn function | all_points = fnplt(cscvn(points))<br><br>Reference matlab literature |
| Built functions | wire_function() | See Table 2 and supplemental information |
| | getInsertedLength()<br>- Calculates the length of wire inside the body.<br>- uses arclength() | getInsertedLength(all_modified_points, points, intersecting_index) |
| | modifyWireLength()<br>- modifies all_points (the interpolated starting_points) to create a variable all_modified_points which will be used in later calculations and plots. | modifyWireLength(all_points, wire_length, nom_length) |
| | plotfunc()<br>- if the user would like to show the 3d plot in the spatial field (by setting the optional variable show_plot to true), this function will plot the modified and unmodified wires, as well as text to the side that shows the length of the wire and the value calculated by the getInsertedLength() function | plotfunc(show_plot,all_points, points,modified_points, wire_length,new_length,distance_inside, outside_length, num_steps) |
| Downloaded functions | arclength()<br>- calculate the length of any sequence of points | arclength(modified_points(1,:), modified_points(2,:), modified_points(3,:)) |

Table 2 The proposed wire_function() variables descriptions.

| | Variable | Description | Examples and Default Values if Applicable |
|---|---|---|---|
| Required Inputs | starting_points | An array with the original control points or nodes that define the spline trajectory of the wire | |
| | ranges | An array of the ranges for each of the starting points in each of the 3 dimensions | |
| | points_changing | The indices of the points in starting_points which will be variable | |
| | s | The users desired electric field | |
| Optional Inputs | num_steps | The desired number of random points to be generated in each range. | If there is a range [-1 2] and num_steps is 3, the function will use 3 random (using the function rand()) values between -1 and 2. The default value for this variable is 5. |
| | nom_length | The desired length of the guidewire | The default value is two meters. |
| | show_graph | A boolean variable which should be set to true if the user wants a histogram of the modified vs. unmodified wire lengths to be displayed (example in fig. 8) | The default value is false. |
| | show_plot | A boolean variable which should be set to true if the user wants a 3d graph in the spatial field to be displayed (example in fig. 3) | The default value is true. |
| | intersecting_index | The index of starting_points which intersects the body, which is used to calculate distance of wire inside and outside of the body. | The default value is 0. |
| | give_warnings | A boolean variable which should be set to true if the user wants notifications when spline points fall outside of s. | The default value is false. |
| | show_electric | A boolean variable which should be set to true if the user wants the graph of the wire in the electric field to be displayed. | The default value is true. |
| Outputs | output_coordinates | A n x m, matrix, where m is the number nodes in the trajectory, and n is the number of coordinates in real space. | Coordinate units are defined by the electric field. |
| | output_modified_wire_lengths | A vector of the final wire trajectory length | The target vector element value is the nominal wire length. |

The wire_function() generates the spline with the native MATLAB library commands cscvn() and fnplt() (fig. 1). The input “starting_points” defines the three coordinates of each initial spline node location, as defined by the user. Wire_function generates a 3-dimensional guidewire path with curves, bends, and loops (fig. 3). In a pop-out figure, the “starting points” can be picked by the user by clicking on an electric field map reference. Generally, the electric field contains enough anatomical structure to be referenced. However, any anatomical reference data - inputted in the variable “s”, which the spline will be graphically laid onto - can be used with the function.

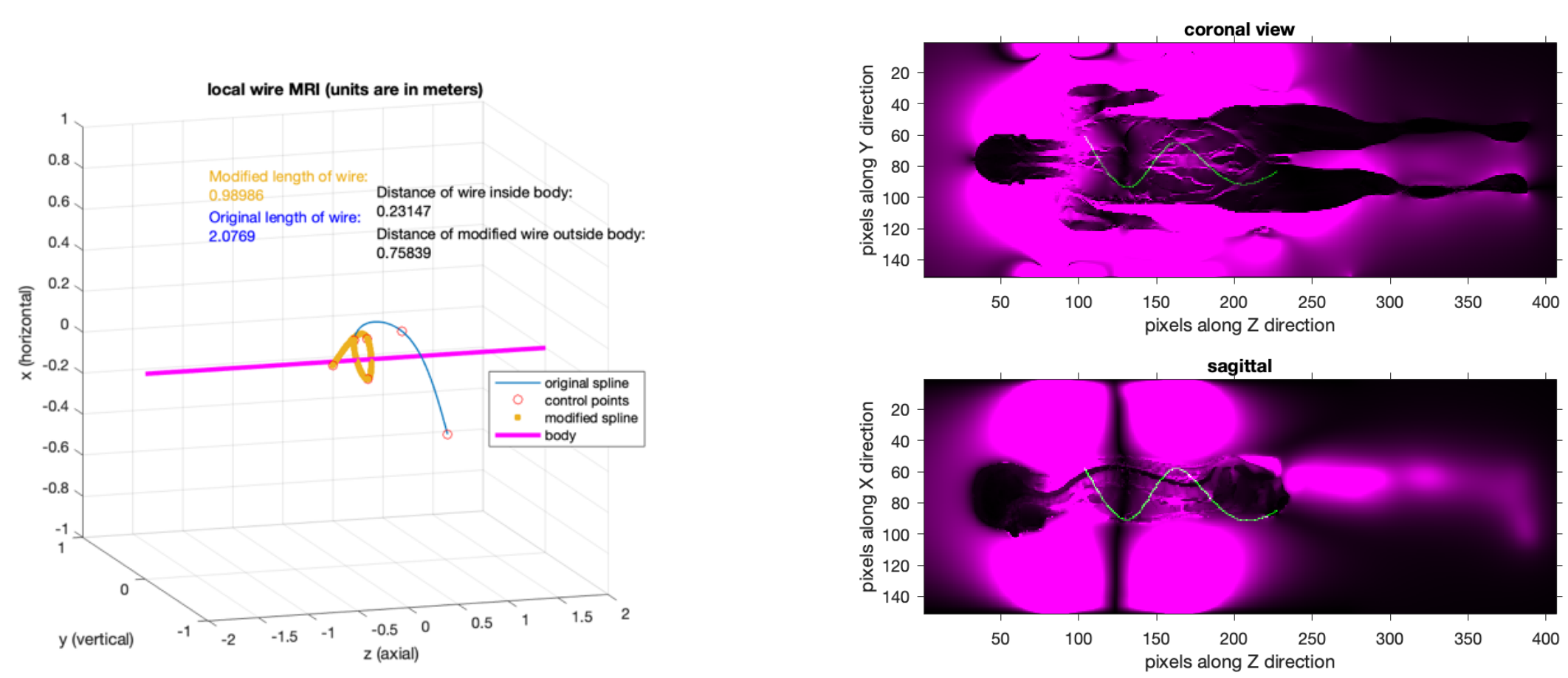

Figure. 3 - Shows two examples of generated wire trajectories. The nominal length (to which the wire length is adjusted) is 1m. Right: shows a curvy trajectory in the electric field. Left: the 3d spatial plot which shows a looped trajectory.

For visualization, the electric field coordinates are displayed in sagittal and coronal views. The guidewire trajectory through the starting points is visualized by creating a matrix of zeros that is equal in length to the dimensions of the electric field. By looping through this field and changing any value with a guidewire coordinate to a 1, the guidewire trajectory is generated. The guidewire trajectory is then fused with the electric field and displayed using the function imshow() (fig. 3, right).

The user may also choose to enter a desired length of the wire in meters; this is an optional input variable called “nom_length”, which prescribes the overall spline length. The function adjusts every simulated trajectory to the nom_length value. If the variable is not entered by the user, the default value is 2 meters. The modified spline length is generated by removing node points from the proximal side of the spline over a “while” loop until the wire length is within ~1cm of the nominal length. There is some variability in the length of the spline that is removed because the distances between points are not fixed. Points are only removed from the spline, not added, because this would require growing the trajectory, which would necessitate extrapolation in an unknown direction.

*Function testing*

The function test reported in this paper is the guidewire length variability and accuracy. The accuracy was calculated as the percent ratio of the final wire length to the nominal wire length value. The function was run for a range of random nominal wire lengths, ranging from 1 meter to 1.8 meters, for 3000 simulated trajectories. The function was also tested for processing speed. We recorded the time it took to run a batch of trajectories, increasing the number of points per trajectory in the same batch until 3000 points were used.

The function was run on a MacBook Air (Apple, Cupertino, California, USA) with an Apple M1 chip, 8GB of memory, and 256GB total storage. An example of values that were tested for the function can be found in the supplemental information.

*Guidewire heating assessment*

To provide a use case example, we present a scenario where 1000 simulated trajectories can be used to study guidewires in the MRI environment. For example, the GTG can be used in research to assess the safety/heating of guidewires when irradiated by either body or local coils. In this example, we compare the predicted scattered electric field at the guidewire tip when excited by a local coil array or a body coil for many random guidewire geometric configurations. Findings have shown that, comparing a body coil to a local coil under worst-case scenario conditions for producing heat on a guidewire tip, local coils reduce the risk of heating. We will briefly describe this research in this section [13] .

The RF fields produced by a body coil and a local coil were compared to determine the different heating effects each coil produces on a given guidewire configuration. The outputs of the GTG were used in the equation for the scattered field at the tip of the guidewire:

$$E_{S,p} = \sum_{l}^{L} H_l E_{T,l} \Delta L$$

Here, *l*, represents a point on the guidewire (GTG output_coordinates). L represents the length of the guidewire (GTG output_modified_wire_lengths). The trajectories of the wires can be used to find $E_T$, the tangential component of the electric field at position *l*. The heating can be predicted with the known incident electric field and knowledge of how that field relates to the scatter field at the guidewire tip [8]. The relationship between the incident electric field and the scattered field at the tip is contained in the electric transfer function $H_l$ (TF) [12] which is unique to the guidewire length and dielectric environment [17].

The scattered electric field produced at the guidewire tip was computed by integrating the product of the tangential component of the electric field produced by the simulated coils along each wire trajectory with the relevant wire TF. The resultant electric field value was squared to approximate a pseudo specific absorption rate (sSAR) and plotted using a violin plot. The mean and median sSAR were determined. The 99th percentile of the sSAR was computed for each coil, and it was considered the worst-case scenario comparison. All data were normalized.

*Transfer function calculation*

The TF was calculated for a straight wire inserted in a dielectric medium leveraging the reciprocal properties of RF transfer functions[18]. The TF was simulated for a 100cm long wire with a 2mm diameter, shielded with 1mm thick plastic, placed in a dielectric gel (permittivity =79, Sigma = 0.47 $Sm^{-1}$) with box dimensions of 500x500x1200$mm^3$. The first 5mm of the tip was exposed by omitting the plastic shielding. The wire was centered in the box, so the guidewire ends were 100mm from the box end. All the TFs were calculated in the EM simulator by exciting the wire with a plane wave at the tip. The pure electric plane wave was generated by overlapping two plane waves with opposing magnetic fields. The resulting electric field was parallel to the wire axis and contained within a wire frame with dimensions 100x100x5$mm^3$ (fig. 4). Multiple insertion depths were simulated: 100cm (fully inserted), 80cm, 70cm, 60cm, 50cm, 40cm.

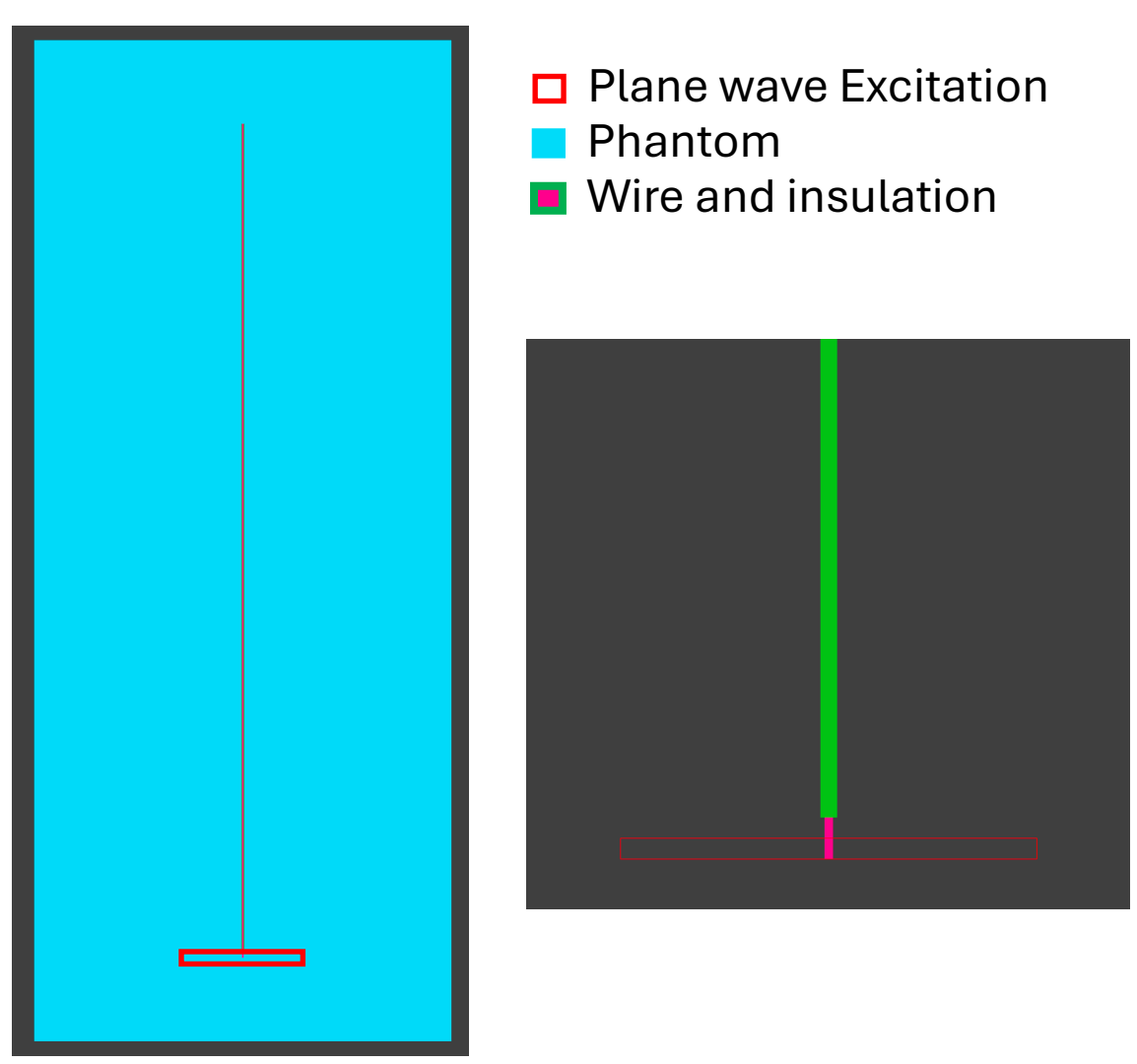

Figure 4 Simulation environment. The blue region is a simulation of the dielectric gel and the red rectangle is the plane wave source. The green represents the wire insulation and the red is the wire material (modeled as PEC, a perfect conductor). The image on the right shows a zoomed view over the wire tip.

## 3. Results

The wire trajectories for a random set of 3000 trajectories from a sample run of the function are shown overlaid onto an electric field map of the Duke phantom (Fig. 5). The individual wire trajectories are shown in green and the electric field map in magenta. The arrow in figure 5 points at the location where the wire enters the body.

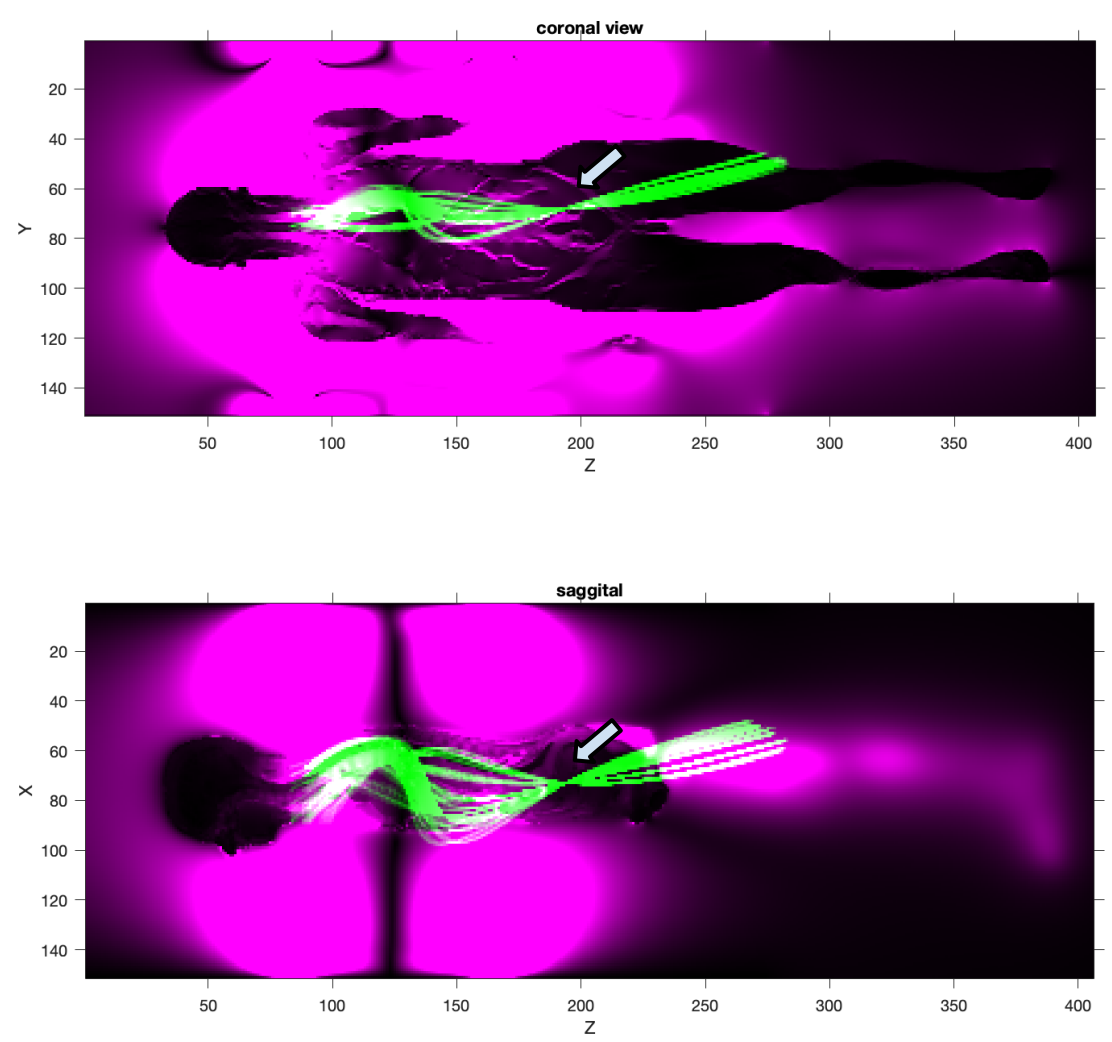

Figure. 5. Shows an overlay of 3000 generated trajectories within an inputted range. This plot is in the electric field. The colormap of the electric field is presented as an anatomical reference, not requiring a quantifiable colorbar.

We tested the number of points on the spline in comparison with length modification accuracy. Figure 6 shows how increasing the number of interpolated points on the spline can affect the accuracy. At 500 points, the accuracy is 99.84%, while at 3000 points, the length accuracy is 99.98%. This difference is insignificant, so we kept the default value for the number of spline points at 500. This is because as the number of interpolated points increases, the time it takes for the function to run also increases (fig. 7). At 500 interpolated points, the average runtime for 3000 trajectories was 11.5 seconds, while at 3000 points the average time is 22 seconds, almost twice as much (fig. 7). A histogram from a sample run of the function with nom_length set to 1.0m is shown in Figure 8.

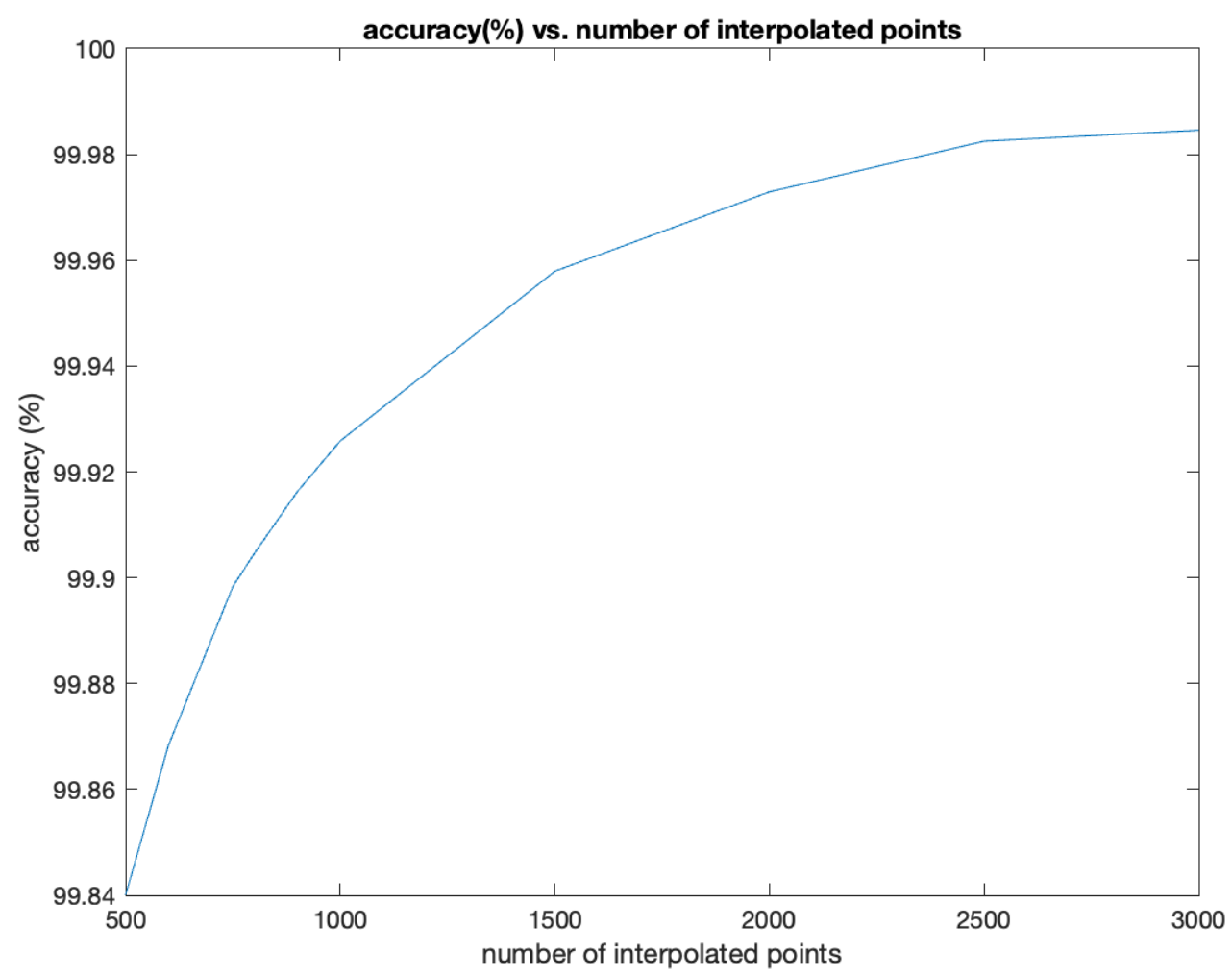

Figure. 6 - The relationship between number of points on the wire (interpolated) and the accuracy of the wire modification. The nominal length is 1.2m.

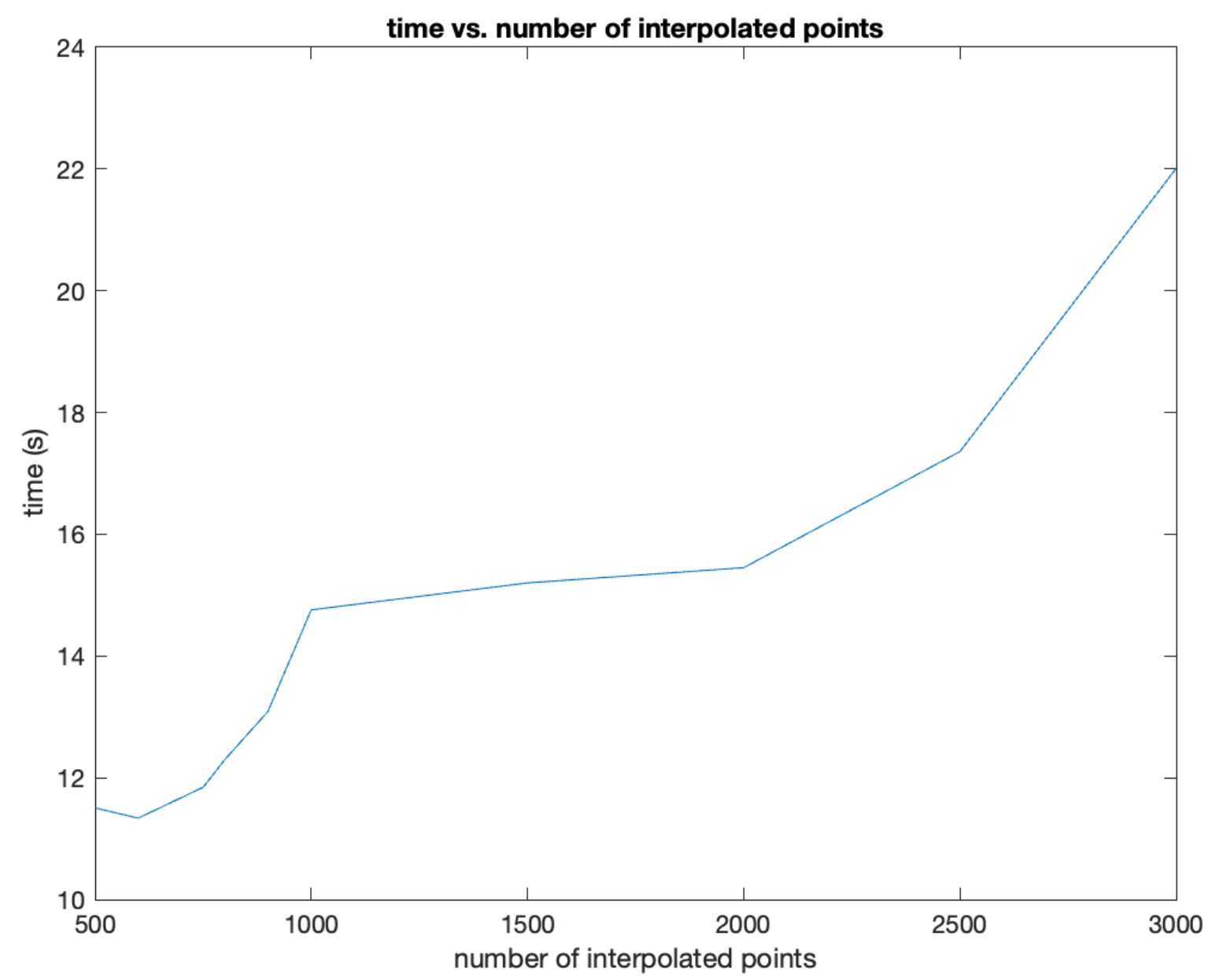

Figure. 7 -  The relationship between the number of points on the wire (interpolated) and the time it takes to run the function (averages) over 3000 trajectory generated.

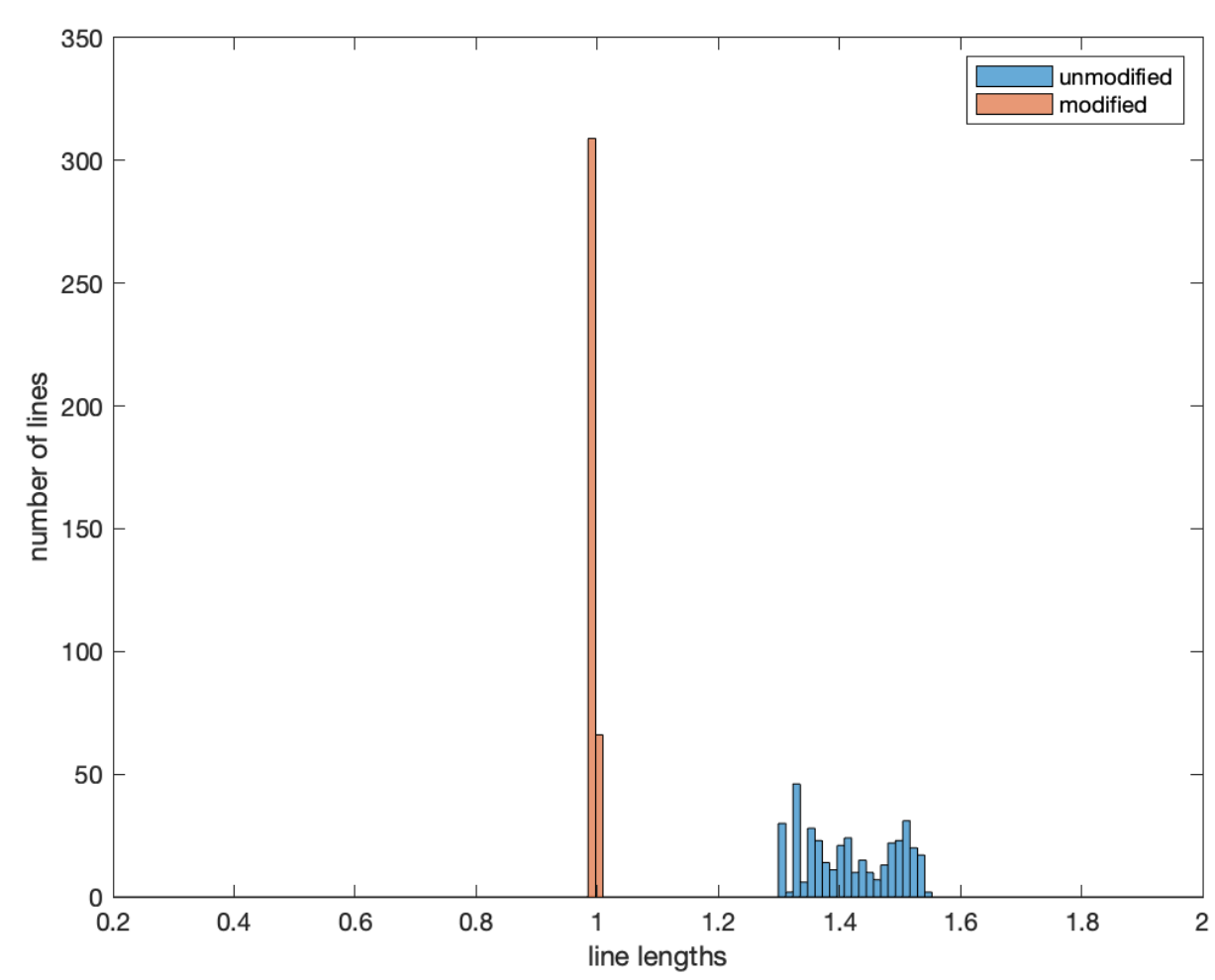

Figure. 8 - Histogram of wire modification from a sample run. Nom_length is 1m.

*Tip heating analysis*

In Figure 9, a violin plot shows the sSAR produced by the local and body coil across all trajectories for the 100cm long wire. The difference between the body coil and local coil array is also shown in Figure 9. The body coil to local coil array ratio, averaged across depths for the 50th , 95th, and 99th percentile values was 1.0, 1.6,  and 2.27, respectively.  The power of the scattered electric field from the body or local coil at the tip is shown in Figure 10 for the worst-case scenario at the relative insertion depth.

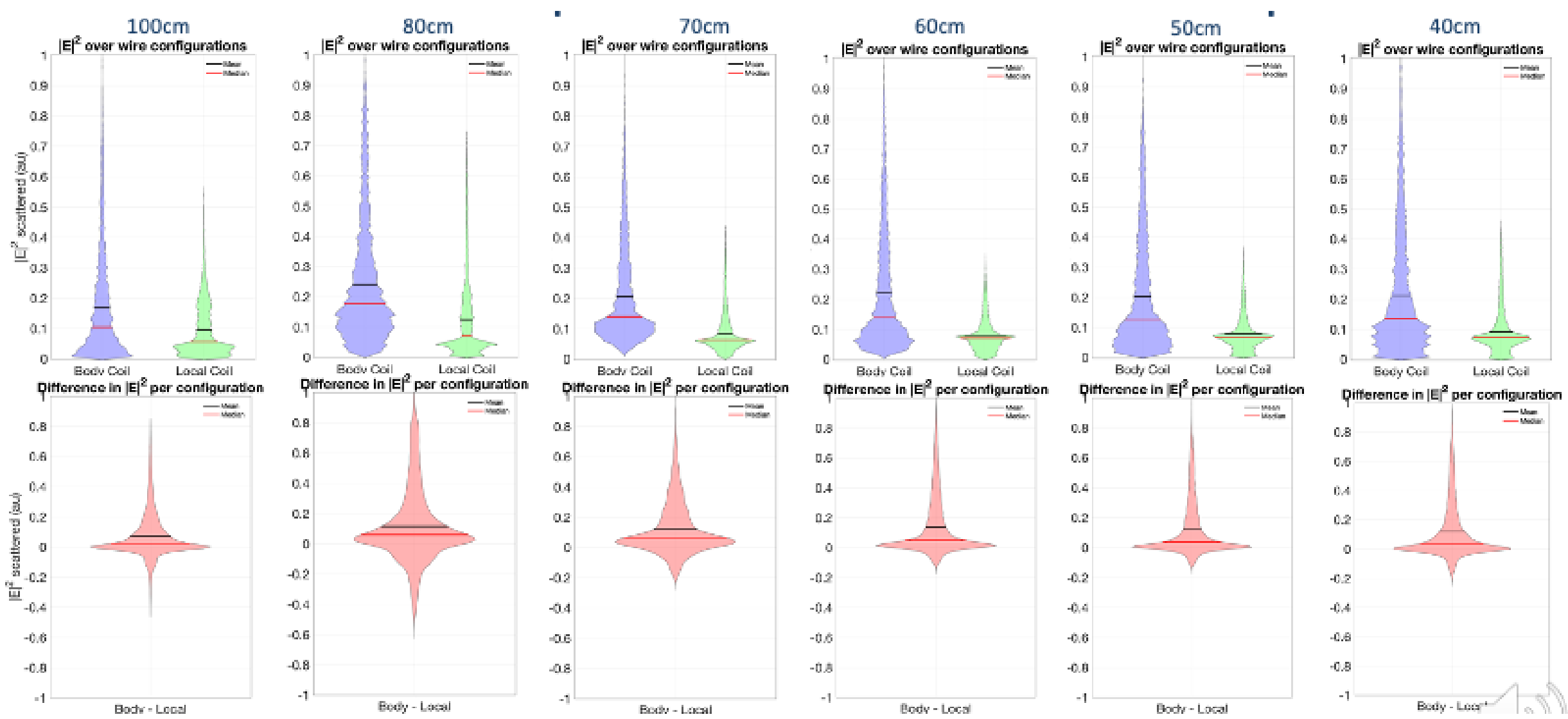

Figure 9 Violin plot of the square of the scattered electric field (sSAR) magnitude for the Body coil and local coil for a 100cm long wire (top) at multiple insertion depths. The difference between the body and local coil (bottom). The data is normalized to maximum value.

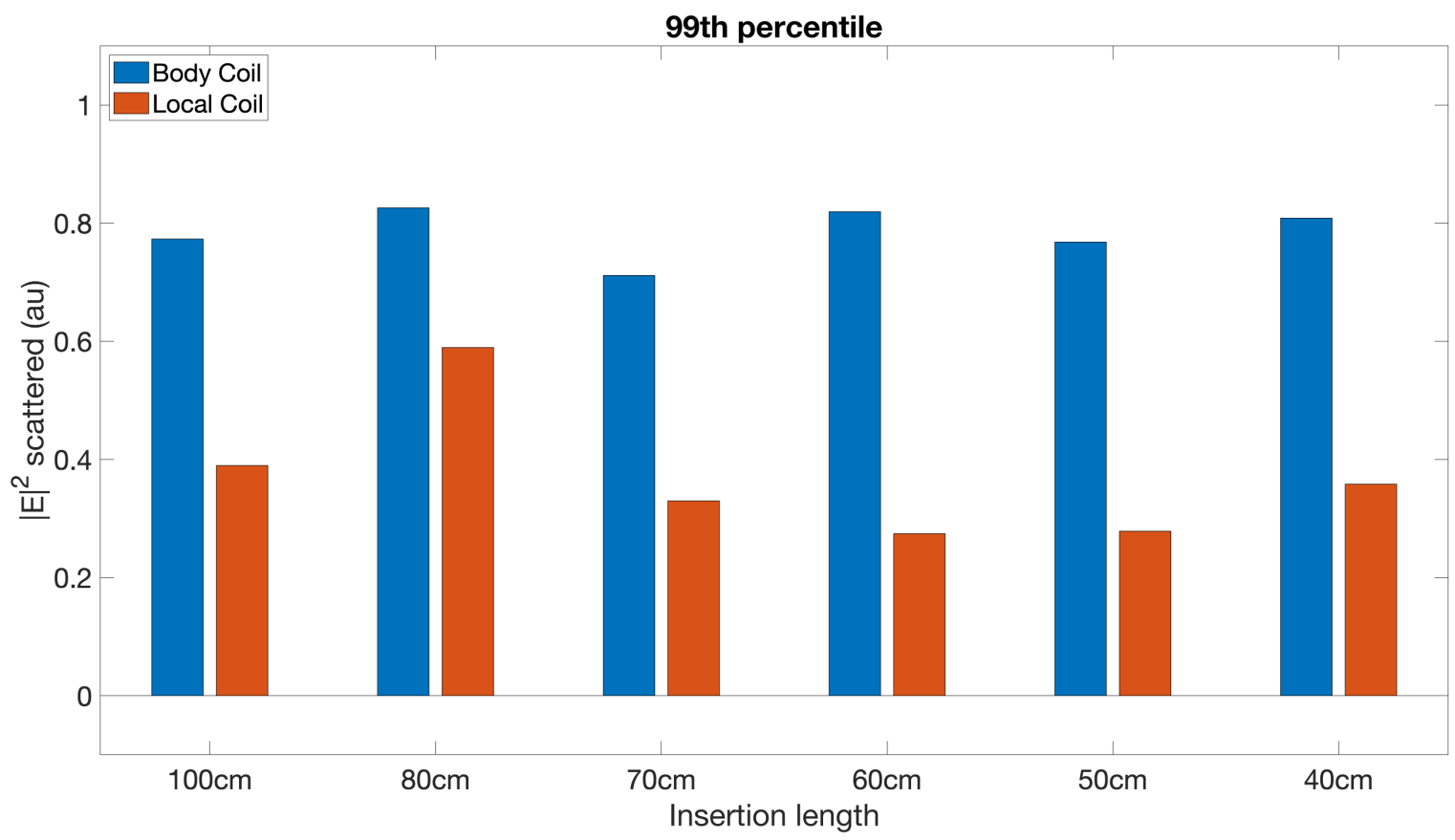

Figure 10 Bar plot of the power (SAR) of the scattered electric field at the wire tip for the body coil and the local coil electric fields at the worst case scenario, $99^{th}$ percentile.

## 4. Discussion

In this work, we developed a tool that will support the simulation of guidewires in MRI environments, making it feasible to randomly model a high number of trajectories. The function is flexible and can accurately model guidewires' lengths and trajectories in a spline-like behavior. This stochastic approach distinguishes the GTG tool from existing simulation methods for interventional MRI heating, which typically evaluate a single, fixed guidewire geometry, such as the fixed-geometry simulations of Zeng et al [9] and Vu et al. [11], or characterise heating from single trajectory experimental measurements, as in Konings et. al. [10] . Because RF-induced heating at a guidewire tip is highly sensitive to the wire's specific length, insertion depth, and path through the body, a heating risk estimate based on one or a few configurations may not reflect the worst case that could occur in practice. By generating thousands of randomized trajectories within clinically plausible ranges, our tool allows researchers to characterize a distribution of heating outcomes and identify outlier or worst-case geometries that a small number of phantom experiments or single-geometry simulations would be unlikely to sample. This capability is particularly relevant to safety-critical applications, such as MR conditional device labeling, where the tail of the risk distribution, rather than the average case, typically determines whether a device or procedure is considered safe.

A key feature of the function is maintaining a nominal length. The accuracy of the length depends on the number of interpolated points, but, as described in the results section, the default number of points (500) creates an error of less than 0.2% (fig 6) without increasing the run time of the function (fig 7). If a user needs an accuracy greater than 99.8%, they can modify the number of interpolated points, increasing the runtime as a side effect. However, given that heating effects are dependent on the resonant length, which is on the order of tens of centimeters at 1.5T, inaccuracies of less than 5 mm are not significant to the heating value. Because the runtime of the function scales with the number of interpolated points and each trajectory can be generated independently, batches of simulations are also well suited to parallelization or distributed computing. This makes it practical to trade the modest runtime cost of higher accuracy against the benefits of running much larger batches, an option that may be valuable for users conducting large-scale statistical studies of guidewire heating risk beyond the 3000-trajectory batches demonstrated here.

One of the features of this tool is that the user can input optional parameters to customize the random trajectory range and display options. Customization of the number of nodes that define the wire can help fix random space to meet anatomical limitations and have more realistic geometries. Users can also customize their display options, including displaying the optional spatial plot (fig. 3) and histogram (fig. 8), which can be useful in making sure that the graphs are specific to the data the user needs, including analysis of the spline's position in space and its accuracy.

Several limitations of the current tool should be noted. First, the trajectory generator enforces a nominal wire length by removing spline points from the proximal end, but does not model the mechanical stiffness or bending resistance of a physical guidewire, so some generated geometries may include curvature that a real device could not assume; incorporating mechanical constraints on curvature or minimum bend radius is a natural extension. Second, the trajectories and RF heating predictions presented here were generated using a single anatomical model (the Duke phantom), a single field strength (1.5T), and a transfer function calibrated for one wire geometry, diameter, and dielectric environment; extending the tool's validation to additional body models, field strengths, and lead geometries would strengthen the generality of its predictions. Third, while the transfer function method allows heating to be estimated without repeated full-wave simulation, the trajectories and predicted heating values generated by the tool have not yet been validated against direct experimental measurements, such as gel phantom testing; such validation would be a valuable next step before the tool is used to support formal safety assessments. Finally, the current implementation and transfer function framework are specific to conductive, metallic guidewires; heating mechanisms and applicable methods for metal-free or hybrid endovascular devices, which are increasingly used clinically, would need to be addressed separately.

Bhidayasiri et al. [19] used an experimental approach to assess heating during deep brain stimulation. In their approach, the electrode tip heating was evaluated with a gel phantom inside an MRI scanner, an in vitro experiment that is not easy to repeat because of cost, time, and materials. Our function allows users to generate data with precision and accuracy quickly. In the future, the function we have created could also be expanded into simulations for deep brain stimulation and assessing the heating associated with it, allowing users to determine many more trajectories in a shorter period. Beyond its use in the present study, making the function and its documentation openly available on GitHub is intended to lower the barrier for other groups to study guidewire and endovascular device heating computationally, without requiring access to gel phantom infrastructure or full electromagnetic simulation expertise. Future extensions beyond deep brain stimulation electrodes could include incorporating patient-specific vascular anatomy in place of a generic phantom as the reference structure for trajectory generation, and adding optional mechanical constraints so that generated trajectories more closely reflect the physical limitations of real devices.

In the use case example, the many guidewire trajectories were useful in evaluating the heating effects from different coil types. The local coil array produces less scattered fields at the guidewire tip on average, 20-34% less, compared to the body coil, when the B1 field in a region of interest is matched. However, our results imply that unsafe conditions can exist in both coil types since there are trajectories for which the local coil array may perform worse (differences are both positive and negative). This initial work helps to provide a general understanding of the complexity of the problem. This finding argues against treating coil type alone as a reliable predictor of heating risk for an individual patient or procedure, since a coil configuration that is safer on average may not be safer for a specific guidewire path. Trajectory-specific or patient-specific risk assessment, rather than blanket assumptions based on coil type, may therefore be warranted when evaluating the safety of a given interventional MRI setup.

## 5. Conclusion

The outputs generated by this function provide users with data about the lengths of the wires generated and the electric field coordinates. These outputs can be used to analyze interactions between RF coils and guidewires, including wire heating and assessing its associated dangers.

The numerous graphs give users visual information, so they know how their spline would look in the electric field and the spatial field. This lessens the possibility of error when calculating wire length and placement. Overall, this function provides users with simulated geometrical data of wire trajectories in the electric and spatial field and is a valuable tool for researchers analyzing the effects of RF-coil/endovascular device interactions and electric scatter at the tip of wires.

## 6. Acknowledgements

This work was supported by the UC Davis Comprehensive Cancer Center through the Paul Calabresi Career Development Award for Clinical Oncology, 5K12CA138464-14, and by the UK Medical Research Council developmental pathway funding MR/N027949/1.